\documentclass[conference]{IEEEtran}
\usepackage{cite}
\usepackage{amsmath,amssymb,amsfonts}
\usepackage{algorithmic}
\usepackage{graphicx}
\usepackage{textcomp}
\usepackage{xcolor}
\usepackage{fancyhdr}
\usepackage[hyphens]{url}
\usepackage[numbers]{natbib}
\usepackage{hyperref}
\usepackage{tabu}
\usepackage{xspace}
\usepackage{enumitem}
\usepackage[nameinlink,capitalise]{cleveref}
\usepackage{pgfplots}
\usepackage{tikz}
\usetikzlibrary{arrows}
\usetikzlibrary{shadows}

\hypersetup{
        colorlinks,
        linkcolor={red!50!black},
        citecolor={blue!80!black},
	urlcolor={blue!80!black}}

\def\BibTeX{{\rm B\kern-.05em{\sc i\kern-.025em b}\kern-.08em
    T\kern-.1667em\lower.7ex\hbox{E}\kern-.125emX}}

\newcommand{\parhead}[1]{\vspace{0pt plus 3pt minus 0pt}\par\noindent\textbf{#1}\hspace{.4em plus .2em minus .2em}}

\crefname{figure}{Figure}{Figures}

\newcommand{\fw}{\textsf{CacheFX}\xspace}
\newcommand{\entropyL}{\textit{Relative Eviction Entropy}\xspace}
\newcommand{\entropyS}{REE\xspace}

\newcommand{\pp}{\textsf{Prime+\allowbreak Probe}\xspace}

\newcommand{\et}{\textsf{Evict+\allowbreak Time}\xspace}

\title{CacheFX: A Framework for Evaluating Cache Security} 
\author{
  \IEEEauthorblockN{Daniel Genkin\IEEEauthorrefmark{1}, William Kosasih\IEEEauthorrefmark{3}, Fangfei Liu\IEEEauthorrefmark{2}, Anna Trikalinou\IEEEauthorrefmark{2}, Thomas Unterluggauer\IEEEauthorrefmark{2}, Yuval Yarom\IEEEauthorrefmark{3}}
\vspace{0.05in}
\IEEEauthorblockA{\IEEEauthorrefmark{1}Georgia Institute of Technology}
\IEEEauthorblockA{\IEEEauthorrefmark{2}Intel Corporation}
\IEEEauthorblockA{\IEEEauthorrefmark{3}University of Adelaide}
}

\begin{document}
\maketitle
\pagestyle{plain}


\begin{abstract}
  Over the last two decades, the danger of sharing resources between programs has
  been repeatedly highlighted.
  Multiple \emph{side-channel attacks}, which seek to exploit shared components for
  leaking information, have been devised,
  mostly targeting shared caching components.
  In response, the research community has proposed multiple cache designs that
  aim at curbing the source of side channels.

  With multiple competing designs, there is a need for assessing
  the level of security against side-channel attacks that each design offers.
  Several metrics have been suggested for performing such evaluations.
  However, these tend to be limited both in terms of the potential adversaries
  they consider and in the applicability of the metric to real-world attacks,
  as opposed to attack techniques. Moreover, all existing metrics implicitly
  assume that a single metric can encompass the nuances of side-channel security.

  In this work we propose \fw, a flexible framework for assessing and
  evaluating the resilience of cache designs to side-channel attacks.
  \fw allows the evaluator to implement various cache designs, victims, and
  attackers, as well as to exercise them for assessing the leakage of information
  via the cache.

  To demonstrate the power of \fw, we implement multiple cache designs and
  replacement algorithms, and devise three evaluation metrics that measure
  different aspects of the caches: 
	(1) the entropy induced by a memory access;
	(2) the complexity of building an eviction set;
	(3) protection against cryptographic attacks;
  Our experiments highlight that different security metrics give different insights
  to designs, making a comprehensive analysis mandatory. 
  For instance, while eviction-set
  building was fastest for randomized skewed caches, these caches featured 
  lower eviction entropy and higher practical attack complexity. 
  Our experiments show that all non-partitioned designs allow for effective cryptographic attacks.
  However, in state-of-the-art secure caches,
  eviction-based attacks are more difficult to mount than occupancy-based attacks,
  highlighting the need to consider the latter in cache design.
\end{abstract}

\section{Introduction}

Memory caches, which store recently accessed memory contents, became a standard feature of mainstream computer processors.
While instrumental to the needs of contemporary computing, sharing caches between multiple untrusted programs can lead to 
undesired information leaks, in that the contents of caches necessarily depend on past computations and by their nature
are intended to enhance the speed of future computations~\cite{GeYCH18}.
By monitoring the timing of memory operations, an attacker can infer the state of the cache
and learn information about the behavior of the victim.
Such \emph{side channels} can result from any of the caches in the 
processor~\cite{OsvikST06,Percival05,AciicmezKS07,Aciicmez07,GrasRBG18,LiuYGHL15,IrazoquiES15}, 
and using such side channels, a malicious actor may seek to infer sensitive information such as 
cryptographic keys~\cite{YaromF14,OsvikST06,BrumleyH09,CabreraGTB19,PereidaB17,MoghimiIE17,GenkinPSYZ20,RonenGGSWY19}, 
user keystrokes and their timing~\cite{RistenpartTSS09,GrussSM15},
address space information~\cite{EvtyushkinPA16,HundWH13,GrasRBBG17},
and others~\cite{ShustermanKHMMO19,OrenKSK15,YanFT20,BrasserMDKCS17}.
The shared use of caches has also been shown to enable efficient 
\emph{covert channels}~\cite{MauriceNHF15,MauriceW0GGBMR17,CockGMH14},
where a malicious Trojan colludes with an attacker to bypass the system's security policy.

The two main types of cache attacks are contention-based attacks~\cite{OsvikST06,Percival05,AciicmezKS07, 
	Aciicmez07,GrasRBG18,LiuYGHL15,IrazoquiES15,YanSGFCT19}, which seek to exploit
the limited storage space in the cache, and
reload-based attacks~\cite{YaromF14,GullaschBK11,GrussSM15,GrasRBBG17}, 
which seek to exploit the attacker's ability to evict memory it
shares with the victim from the cache.
Because reload-based attacks rely on shared memory, preventing memory from being shared across
security domains can be an effective countermeasure.

Many cache designs have been suggested to address contention-based cache attacks:
partitioned caches aim to prevent contention~\cite{WangL07,DomnitserJLAP12} while 
randomized caches~\cite{LiuWML16,WernerUG0GM19,Qureshi18,Qureshi19,TanZBR20} aim
to introduce noise and prevent the attacker from analyzing the side-channel signal.
Randomized caches often try to prevent the attacker from mapping addresses 
to predictable cache line indexes, a step that is considered essential for the attack.
Finally, some proposals try to prevent cross-core attacks by tweaking the inclusion
properties of shared cache levels in modern processors~\cite{YanWFT19,GreenLZIHE17,KayaalpKEEAPJ17,YanGST17}.

With multiple proposals for protecting against contention, processor vendors need some
method of assessing the security these proposals provide.
Several approaches for evaluating secure caches have been 
suggested~\cite{KopfMO12, DoychevFKMR13, Ghasempouri2020, ZhangL14, HeL17, Deng0S20, DengXS19, WangZWM19, ZhangLCL13, DemmeMWS12, DemmeMWS13,  CockGMH14, GeYH18, GeYCH19}. 
For instance, \cite{KopfMO12, DoychevFKMR13, Ghasempouri2020} use formal methods and model checking to determine cache leakage, \cite{WangZWM19, HeL17} model cache attacks to obtain attack success probabilities, and \cite{Deng0S20, DengXS19} use a three-step attack model to exhaustively test for vulnerable attack patterns and apply it to various Arm devices~\cite{DengMXKS21}.
However, all of those suffer from some limitations as they only work with
simple cache models, focus on theoretical analysis, cannot be automated, or do
not cover the full range of cache attacks. In addition, to strengthen
confidence in the security of cache designs, the evaluation of multiple metrics
is mandatory. Thus, in this work, we investigate the following question.

\smallskip
\begin{center}
\emph{How can we evaluate the security that cache designs offer against contention-based
cache attacks?}
\end{center}
\smallskip

\subsection{Our Contribution.}
To address this question, this paper presents \fw, a framework for evaluating the security
of caches. \fw provides an interface for emulating the operation of cache designs with different victims and attackers, and measuring the leakage for each combination of attacker, victim, and cache design.

\parhead{Evaluating Cache Designs.} We implement nine cache designs, including traditional fully-associative and set-associative caches,  PLCache~\cite{WangL07},  Newcache~\cite{LiuWML16},  PhantomCache~\cite{TanZBR20}, 
ScatterCache~\cite{WernerUG0GM19}, way-partitioned caches~\cite{DomnitserJLAP12},  and the two variants of CEASER~\cite{Qureshi18,Qureshi19}.
For caches that do not stipulate a replacement policy, we support four replacement
algorithms: random replacement, least recently used (LRU)~\cite{Denning68}, and two
variants of pseudo-LRU. We then use \fw to evaluate these caches using three metrics:

\begin{itemize}[nosep,leftmargin=*]
  \item \parhead{\entropyL.} 
    The \entropyL (\entropyS) is a new metric we propose to measure the
    information leakage from a single victim access via the cache side channel. 
    For this case, we use an unrealistic attacker that can set the cache to a known
    configuration and accurately observe the cache state. We combine this attacker 
    with a victim that accesses a single cache line only and then calculate the amount of
    information that the attacker receives from observing which cache line 
    the victim evicted. 
  \item \parhead{Measuring Eviction-Set Creation.}
    Most cache-based side channel attacks require the attacker to find a
    minimal set of addresses such that accessing them results in evicting a
    specific victim line from the cache.  Our second metric measures the
    difficulty for an attacker to find such sets. Here, we have implemented
    several eviction-set building strategies, such as the Single Holdout Method
    (SHM)~\cite{Qureshi19}, Group Elimination Method (GEM)~\cite{Qureshi19} and
    Prime+Prune+Probe (PPP)~\cite{Purnal19}.  For each cache design, we
    evaluate each strategy in terms of the number of memory accesses required,
    valid addresses found that collide with the victim, and eviction success
    rate. 

  \item \parhead{Cryptographic Attack.}
    Cryptographic attacks evaluate the protection that the cache provides for vulnerable cryptographic code.
    Our victims perform cryptographic functions using implementations known to
    be vulnerable.
    We use the victims to encrypt data with one of two keys,
    and task the attacker with distinguishing between the keys.
    We evaluate both traditional attacks that aim to exploit eviction sets,
    and occupancy-based attacks.
    For the former, we identify a cache line whose access probability depends on the key
    and provide the attacker with an eviction set for the line.
    For the latter, the attacker accesses a cache-size buffer, with the aim of contending with the victim on cache capacity.
    In both attacks, we use the number of encryptions the attacker needs to observe to distinguish between
    the keys as the security metric.

\end{itemize}

\parhead{Multiple Metrics.}
We evaluate all three metrics on all nine cache designs under multiple parameters.
We find that different metrics highlight different aspects of the
caches.
In particular, we find that building eviction sets is faster in skewed caches such
as ScatterCache and CEASER-S, than other randomized caches, such as fully associative caches 
or PhantomCache.
Faster eviction-set construction reduces the effort required for mounting attacks.
At the same time, our experiments show that, with the right parameters, skewed caches are not
less secure when it comes to cryptographic attacks.

\parhead{Evaluating Cryptographic Attacks.}
Past metrics measure the leakage capacity using synthesized adversaries.
In contrast, two of our metrics use cryptographic attacks, providing more realistic insights on an attacker's capabilities.
We find that the security against cryptographic attackers depends not only on the design,
but also on other parameters, such as the replacement policy and the cache associativity.
We also show that \emph{all} non-partitioned caches are 
vulnerable to both eviction-set and occupancy attacks.
We note that partitioned caches are not necessarily a solution for cache-based
side channel attacks, because they can only support a limited number of concurrent
processes and time-sharing them raises a potential for further leakage~\cite{GeYH18, GeYCH19}.

\parhead{Comparing Attacks.}
We further find that \fw enables us to compare attack strategies across the
various cache designs and parameters.
Thus, we find that of the three eviction-set construction strategies, PPP tends to produce
more accurate results than the other approaches.
Similarly, for cryptographic attacks, we find that most secure cache designs offer protection against eviction-set attacks.
However, cache-occupancy attacks are left unconsidered and for highly secure designs
occupancy attacks are no less effective than eviction-set attacks.

\parhead{Summary of Contributions.}
In summary, this paper makes the following contributions
\begin{itemize}[nosep,leftmargin=1em]
  \item We describe \fw, a framework for evaluating the resilience of 
    caches to microarchitectural side-channel attacks. 
  \item We provide implementations of nine cache designs and four cache replacement
    algorithms, and evaluate these using three sample metrics.
  \item 
    We show that no single metric is sufficient to fully characterize
    the security of cache designs.
  \item We find that all non-partitioned cache designs leak enough information to allow 
    cryptographic attacks.
  \item We note that cache-occupancy attacks, which are not considered by many designs, do pose a threat and with some cache designs can be more efficient than eviction-set attacks.
\end{itemize}
\fw is available at \url{https://github.com/0xADE1A1DE/CacheFX}.

\parhead{Roadmap}
The rest of this document is organized as follows:
In \autoref{s:background}, we provide the required background on caches, cache attacks, and secure cache designs.
\autoref{s:problem} presents the problem that \fw aims to solve.
We present \fw in \autoref{s:design}.
\autoref{s:eval} presents the evaluation of the caches under the three selected metrics.
We then discuss the threats to validity in \autoref{s:validity} and the related work in \autoref{s:related}.

\section{Background}\label{s:background}

\subsection{Cache Attacks}
Modern processors use an array of \emph{caches} to speed up accesses to memory by 
exploiting spatial and temporal locality. Caches are architecturally 
transparent---whether a specific piece of data is cached does not affect 
the architectural behavior of a program. It does, however, affect the performance of programs,
thus monitoring program performance can reveal information about the
state of the cache.

\citet{TsunooTMM02} were the first to demonstrate that
this information can be used to recover secret cryptographic keys.
Early attacks focused on the L1 data cache~\cite{TsunooSSSM03,OsvikST06,Percival05},
but attacks targeting other caches soon emerged~\cite{AciicmezKS07,YanSGFCT19,
Aciicmez07, GrasRBG18,LiuYGHL15,IrazoquiES15, GullaschBK11, YaromF14}.
Cache attacks target symmetric cryptography~\cite{IrazoquiES15,TsunooTMM02,TsunooSSSM03,MoghimiIE17,GenkinPSYZ20,OsvikST06, GullaschBK11},
public-key schemes~\cite{YaromF14, AciicmezKS07, Percival05,
GrasRBG18,LiuYGHL15, PereidaB17, RonenGGSWY19,DallMEGHMY18},
Post-quantum cryptography~\cite{BruinderinkHLY16,PesslBY17}, and
non cryptographic software~\cite{GrasRBBG17,GrussSM15,HundWH13,EvtyushkinPA16,
ShustermanKHMMO19, OrenKSK15,YanFT20,BrasserMDKCS17}.
At a high level, cache attacks can be divided into two main groups.

\parhead{Reload-Based Attacks}  
monitor accesses to a shared memory address~\cite{GrussSM15,GullaschBK11, YaromF14}.
The attack first evicts the data from the cache either via
a dedicated instruction~\cite{GullaschBK11, YaromF14} or by forcing
contention on the cache set containing the data~\cite{GrussSM15}.
The attacker then waits a bit and measures the time to access the 
previously evicted data.
If while waiting the victim accesses the data, the data will be cached and
the attacker's access be fast. 
As not sharing memory across domains can be an effective
mitigation, this attack type is not the main target of this work.

\parhead{Contention-Based Attacks},
which seek to 
exploit the limited storage in the cache,
and in particular in each of the cache sets~\cite{OsvikST06,Percival05, AciicmezKS07, 
	Aciicmez07, GrasRBG18,LiuYGHL15,IrazoquiES15,YanSGFCT19},
are the main focus of this work. 
The most common contention-based attack technique is \pp, where
the attacker first primes the cache by filling some or all of the cache sets with
their data and, after letting the victim some time to execute, measures the time
to access the cached data.
A slow access indicates that the data is no longer cached, suggesting eviction
from a cache set due to victim activity.
Variants of the attack avoid using timing information by 
relying on  performance counters~\cite{BhattacharyaM15,UhsadelGV08,BrasserMDKCS17} 
or transaction aborts~\cite{DisselkoenKPT17}
for contention detection.
Another contention-based attack is \et~\cite{OsvikST06,GrasRBG18,vanSchaikRGBG17,JainBR19}, 
where the attacker evicts data from the
cache before measuring the execution time of a victim.  
The victim's execution time will be longer in the case that the evicted data
is used by the victim, revealing information on cache sets that the victim uses.

\parhead{Other Types of Cache Attacks.}
Some cache attacks do not fit into either of the two groups.
Such  attacks seek to exploit implementation aspects of the cache, 
such as port contention~\cite{YaromGH17}, 
cache flushing time~\cite{GrussMWM16},
replacement policies~\cite{XiongS20,VilaAGKR20,PurnalTV21},
cache inspection operations~\cite{Brumley15}, or variations in power
consumption based on caching information~\cite{Page02}.
Due to their specific requirements, such attacks are outside the scope of this work.

\parhead{Eviction-Set Construction.}
For many of the attacks mentioned above, the attacker needs to be able
to repeatedly evict specific contents from the cache.  
Typically, attackers achieve this by constructing an \emph{eviction set},
which consists of memory locations that all map to the same cache set
as the data to evict.
When mapping information for the cache is available, constructing an
eviction set tends to be straightforward.
However,  when  the mapping function is undocumented or when the information it uses for
indexing the cache is not available to the attacker, additional techniques are required to recover the missing information. Tackling this problem, past research shows how to reverse-engineer undocumented mapping 
functions~\cite{McCalpin21,MauriceSNHF15,InciGIES16,YaromGLLH15,GrasRBG18},
and how to construct eviction sets in the absence of physical address
information~\cite{LiuYGHL15,VilaKM19}.

\subsection{Secure Caches}
Several proposed cache designs aim to mitigate contention-based attacks.
Their mitigation strategies are either based on partitioning~\cite{WangL07,DomnitserJLAP12} or randomization~\cite{LiuWML16,WernerUG0GM19,Qureshi18,Qureshi19}.

\parhead{Partitioned Caches}
 Way-partitioned caches~\cite{DomnitserJLAP12} enforce a strong
partitioning between security domains by letting each security domain use a different subset of 
the cache ways. Hence, domains not sharing cache ways will not see any interference. 
Alternatively, {Partition-Locked (PL)~\cite{WangL07}}  caches share the whole cache 
among all security domains, but offer to pin cache lines in the cache. These pinned lines
cannot be evicted by other security domains, preventing contention-based attacks. However,
aggressive pinning can starve other domains and severely degrade their performance.

\parhead{CEASER.} The CEASER cache~\cite{Qureshi18} is based on an ordinary set-associative cache and 
uses encryption to randomize the mapping of addresses to cache
sets. As a result, attackers need to first profile the victim's accesses of interest
to find a suitable eviction set before they can perform contention-based attacks. To
limit the attacker's time for finding such eviction set, CEASER regularly changes
the encryption key. 
However, in this work we only measure information leakage in each key epoch (i.e., during the time period the cache uses the same key) and do not model re-keying. 
This allows assessing the 
security of pure cache-set randomization as it is needed to appropriately tune the 
re-keying interval for long-term security. 

\parhead{CEASER-S and ScatterCache.} As eviction set building techniques improved~\cite{VilaKM19,Qureshi19},
CEASER was required to increase its key refreshing rate to maintain security, resulting in increased overheads.
CEASER-S~\cite{Qureshi19} and ScatterCache~\cite{WernerUG0GM19} thus use a skewed cache~\cite{Seznec93}
to improve cache randomization. These skewed caches split
the cache into partitions along its ways and use a different key to encrypt the 
address to index into each partition. The number of partitions can vary between 1 (CEASER)
and the number of ways (ScatterCache) and allows to control the degree of randomization. As in CEASER, 
we avoid re-keying CEASER-S and ScatterCache to assess 
the security of pure cache randomization with skewing. 

\parhead{PhantomCache.} PhantomCache~\cite{TanZBR20} builds upon set-associative caches
and maps each address to multiple sets via multiple hash functions, i.e., it looks 
in multiple sets for a cache hit. If there is a miss, PhatomCache randomly selects one of the 
sets it maps to and inserts the cache line into the chosen set. The number of cache sets 
looked up in parallel determines the degree of randomization and the cost of lookup. 
As before,
we evaluate PhantomCache without re-keying. 

\parhead{NewCache.} 
Rather than randomizing the cache mapping,
NewCache~\cite{LiuWML16} is
a more efficient implementation variant of a fully associative cache. 
NewCache allows every cache line to be stored in any of the physical lines of
the cache. Compared to a standard associative design, it optimizes power using
a two-step look-up procedure: For a cache that can hold $2^n$ physical cache
lines, NewCache first looks up $n+k$ index bits of the cache line address in a
$2^n$-element Content-Addressable Memory (CAM), which has a 1:1 mapping to the
actual cache lines. Only if these $n+k$ bits match, this \emph{index hit} is
secondly followed by checking the tag for the respective entry. If the
\emph{tag hits}, the cache line is found and returned. If there is a \emph{tag
miss} for the same security domain in the second step, the tag and cache line
are simply replaced. If there is a \emph{index miss}, any of the $2^n$ cache
lines in the cache is randomly replaced. While for large $k$ NewCache resembles
a traditional fully associative cache, a smaller $k$ significantly reduces
power and implementation cost. 

\parhead{Secure Cache Hierarchies.}
For cross-core attacks, attackers must be able to evict
data not only from the caches they use, but also from caches at the victim's core.
Most cross-core attacks rely on an inclusive LLC,
which ensures that the contents of the shared LLC is a superset
of the contents of all private caches.
Inclusiveness guarantees that when data is evicted from the LLC,
it is also evicted from all core-private caches.
If the LLC is not inclusive, evictions from it do not necessarily
translate to evictions from core-private caches, and may hamper cache attacks.
\citet{YanSGFCT19} use a similar property of cache directories in non-inclusive caches.
Several cache designs and features~\cite{YanWFT19,GreenLZIHE17,Panda19} that
prevent cross-core eviction are outside the scope for this work.

\section{Problem Description}\label{s:problem}
With the abundance of secure cache designs, there is a clear need for systematically evaluating 
the security of caches to ensure that emerging cache architectures deliver the promised
protection. Tackling this task, previous works~\cite{DengXS19, Deng0S20,
DomnitserAP10, DemmeMWS12, DemmeMWS13, DemmeS14, HeL17, WangZWM19, ZhangLCL13,
ZhangL14, CockGMH14} have suggested several metrics.  However,
all of these tend to suffer from some limitations to their practicality.
For example, measuring the amount of information that can be transferred
via a cache side channel~\cite{CockGMH14} or the correlation between
a specific victim's activity and attacker observation~\cite{DemmeMWS12} may not
translate easily to cryptographic attack scenarios.
Possibly the most common limitation of these approaches is the attempt to provide
a single metric that somehow represents the security of the cache.

\parhead{A General Cache Evaluation Framework.}
Instead of focusing on a single metric, this work 
proposes \fw, a framework for evaluating 
the security of cache designs.
The main design aim of \fw is flexibility. That is,
\fw is extensible and allows evaluating various combinations
of victims, cache designs, and attack strategies.
As a proof of \fw's generality, this paper implements 
and evaluates three security metrics on nine different cache designs.

\parhead{A Leakage Upper Bound.}
While we try to evaluate in realistic scenarios, \fw aims to provide
an upper bound on the amount of leakage an attacker can obtain from a
cache design. We thus assume an attacker who has significant
control over the victim and is tightly synchronized with the victim's
execution.  We further assume that the attacker has access to victim's
memory layout and code and thus knows the position of
``interesting'' data in the cache (e.g., cache lines containing secrets). 
This allows the attacker to craft inputs to the victim that may cause specific cache
footprints. Except as required by the cache design,  we assume a
noise-free environment without any system activity besides the attacker and
victim.  

We note that using such strong assumptions allows \fw to properly evaluate the
security offered by the cache design, as opposed to being misled by security
guarantees stemming from noise from other components. Finally, we note 
that previous works have demonstrated that attackers can find interesting cache lines~\cite{LiuYGHL15} and  overcome
noise~\cite{CockGMH14}.

\section{\fw Design}\label{s:design}

\begin{figure}
 \begin{center}
   {\footnotesize
   \begin{tikzpicture}
     \tikzstyle{textbox} = [drop shadow={opacity=.8,shadow yshift=-1ex, shadow xshift=1ex}, fill=white,  rectangle, minimum height=.8cm, minimum width=1.8cm, text centered, draw=black]
     \node [textbox] (AM) at (0,0) {Attack Model};
     \node [textbox] (MH) at (3.2,0) {Memory Handle};
     \node [textbox] (CM) at (6.4,0) {Cache Model};
     \draw [-stealth', transform canvas={yshift=0.12cm}] (AM) -- (MH) ;
     \node at (1.6,.3) {rd offset};
     \draw [-stealth', transform canvas={yshift=0.12cm}] (MH) -- (CM) ;
     \node at (4.9,.3) {rd offset};
     \draw [densely dotted, -stealth', transform canvas={yshift=-0.12cm}] (MH) -- (AM) ;
     \node at (1.6,-.3) {hit/miss};
     \draw [densely dotted, -stealth', transform canvas={yshift=-0.12cm}] (CM) -- (MH) ;
     \node at (4.9,-.3) {hit/miss};
   \end{tikzpicture}
   }

 \end{center}
\caption{\fw design overview.}
\label{fig:cachefx_design_overview}
\end{figure} 

As mentioned above, \fw is designed to provide an easily extendable framework
for (1) evaluating the security of emerging cache designs and (2) the
applicability and complexity of new attack strategies to both deployed and
emerging caches. To facilitate these goals, \fw is split into three major
components as depicted in \autoref{fig:cachefx_design_overview}. First, the
\texttt{attack model} provides a set of interfaces and their implementations to
model different attack and security evaluation strategies. The \texttt{attack
models} use a \texttt{memory handle} to request reads, writes, and
cache line invalidations to the memory system by specifying a certain offset
into a memory region that is associated with the \texttt{memory handle}. The
\texttt{memory handle} translates the requests to cache line addresses and
queries the \texttt{cache model} correspondingly. The \texttt{cache model}
returns whether the request hit or missed in the cache via the \texttt{memory
handle} to the \texttt{attack model}, which then proceeds with the attack
accordingly. Finally, the \texttt{cache model} provides a generic cache
interface allowing for  various different cache implementations.

\subsection{Cache Model}\label{sec:cache-model}

\fw's \texttt{cache model} offers a generic cache interface that the \texttt{memory
handle} and the \texttt{attacker model} can use to issue read, write, and
invalidation requests to the cache under test. For each of these requests, the
cache responds with whether the request hit or missed. This 
indication removes the need to distinguish between hits and
misses using (potentially noisy) timing measurements, providing an
upper bound on the amount of leakage available to the attacker and consequently
lower bounding the attack's complexity in practice. 

\parhead{Supported Cache Designs.}The \texttt{cache model} currently provides
multiple implementations of both state-of-the-art and novel security-oriented
cache designs: fully associative cache, set-associative cache, way-partitioned cache, partition-locked cache,  CEASER and
CEASER-S~\cite{VilaKM19,Qureshi19},  ScatterCache~\cite{WernerUG0GM19}, 
NewCache~\cite{LiuWML16}, and PhantomCache~\cite{TanZBR20}. These cache
implementations are parameterized by the number of sets, ways, replacement
policy, and cache-specific parameters.
In particular, unless a cache design mandates a
specific replacement policy, all the implementations support
LRU, Bit-PLRU, Tree-PLRU, and random replacement. 

\parhead{Statistics Generation.} The abstract cache model automatically tracks
the number of cache hits and misses for the accesses made by each security domain.
In addition, the cache model can return the evicted address, if a cache accessed causes an eviction.
While attackers usually do not have direct access
to such information, providing the address allows us to apply novel and efficient
techniques, such as the \entropyL (\entropyS) in \cref{sec:evaluation_entropy},
for analyzing the security of cache designs.  

\subsection{Attack Model}
\fw's \texttt{attack model} implements the actual adversarial strategy 
and evaluates the cache design under test. Currently, \fw supports three
security evaluation strategies:

\parhead{The Attacker.}
\fw allows to model synchronized pairs of victims and attackers, aiming to
evaluate the security of cache designs with respect to realistic attacks, such
as cache attacks against cryptographic block ciphers.

\parhead{Information Leakage Assessment.}
\fw also supports entropy-based security metrics that quantify information leakage
during cache attacks (e.g., mutual information analysis). Most noteworthy, \fw
implements a novel technique for evaluating information leakage in cache
designs via the \entropyS, by efficiently analyzing the statistical properties
of a cache's cross-domain eviction behavior.

\parhead{Eviction-set Profiling.} 
\fw provides an environment that allows for the evaluation of
strategies to construct eviction sets for different cache designs.

\parhead{Experiment Randomization and Automation.}
\fw allows to conduct each of these experiments multiple times with randomized address ranges to automatically obtain statistical data like maximum, minimum, etc. \fw hereby collects data such as cache statistics and attack success rates.  

\section{Evaluation}\label{s:eval}
Recognizing that no single metric is sufficient for measuring the
resilience of caches to side-channel attacks, we evaluate emerging cache designs w.r.t.\ multiple metrics using our framework \fw.   
First, the \entropyL (\entropyS) metric measures the amount of information 
(in bits) that an attacker can deduce following a single memory access performed by the
victim. 
Our second metric measures the complexity of creating eviction sets in randomized caches. 
Our third metric measures the complexity of performing cache attacks on cryptographic implementations.
It evaluates both traditional attacks that seek to exploit eviction sets
and cache-occupancy attacks~\cite{ShustermanKHMMO19,CockGMH14}, which do not require eviction sets.

We now discuss each metric in detail 
and compare different designs according to each of the measurement metrics.   

\subsection{Relative Eviction Entropy}
\label{sec:evaluation_entropy}
In this section we introduce our \entropyL (\entropyS) technique for effectively measuring the amount of information available to an attacker following a single memory access performed by the victim. We begin by observing that traditional mutual information analysis techniques~\cite{CockGMH14,ZhangL14} achieve such estimation for general side channels by computing a 2-dimensional joint probability distribution, which describes the likelihood of
 each victim activity (side channel input) to  be mapped to an effect observable by an attacker (side channel output). For the case of caches, this implies that for any address $i$ accessed by the victim, and for all cached addresses $a$, we need to compute $p_e(a,i)$ which is the probability that $a$ is evicted from the cache assuming that the victim accesses address $i$. Finally, we note that mutual information techniques typically measure average leakage across accesses, and thus do not capture the worst-case leakage. 
 
\parhead{Avoiding Quadratic Overheads.} 
In order to avoid the quadratic overhead associated with computing the 2-dimensional joint probability distribution, we start by observing that natural cache designs typically
do not have different eviction behavior between cache line addresses, and instead 
 use the same replacement policy constantly across all cache lines. In addition to simplifying cache designs, this property implies that  all cache line addresses exhibit the same leakage behavior. Leveraging this fact,  
 we can thus fix an arbitrary address $i$ to be accessed by the victim, and simply sample $p_e(a,i)$ for all other addresses $a$.  This allows us to avoid the need to iterate over all possible values of $i$, thus making the evaluation time of our metric linear in the size of the victim and attacker address spaces. When the value of $i$ is fixed and clear from the context, we will simply omit $i$ from the notation.  

\parhead{Quantifying Information Leakage.}
Next, in order to capture the amount of leakage available to the attacker (in bits), we start from the intuition that fully associative caches utilizing a random replacement policy leak the least amount of information among all cache designs that share cache lines between security domains, i.e., without consideration of partitioned caches. We argue that this assumption is reasonable, since fully associative caches with uniformly-random replacement only leak whether an address $a$ was evicted or not, and do not reveal any information about which address $i$ accessed by the victim caused the eviction of $a$. 
To evaluate leakage of new cache designs, we thus measure the \entropyS as the statistical distance (in bits) of the eviction behavior of the tested cache design from the eviction behavior of ideal fully associative caches with random replacement.

\parhead{Computing Relative Eviction Entropy.} More specifically, our strategy for computing a cache design's \entropyS is as follows. First, we allocate a chunk of memory in the adversary's address space, typically a small multiple of the cache size. We denote the set of cache line addresses within that adversary's memory as $a \in [0,...,N-1]$. Second, for a single victim access to some fixed address $i$, we estimate the eviction probability  $p_e(a)$ for each cache line $a\in [0,...,N-1]$ in the adversary's memory, using our implementation of the cache design under test. The distribution $p_e(a)$ will reflect the cache's placement policy: e.g., if the single victim access can evict every adversary address $a$, as in a fully associative cache with random replacement, $p_e(a)$ will be uniform among all adversary addresses $a$. If the single victim access can only evict adversary addresses $a$ mapping to the same cache set in a set-associative cache, $p_e(a)$ will be uniform among those addresses $a$ mapping to the same set as the victim address $i$ and zero otherwise. The reference eviction distribution of a fully associative cache with random replacement is set to $p_u(a) = 1/N$ for all addresses $a$, reflecting that every adversary address is equally likely to get evicted. 
Finally, we compute the \entropyS as the statistical distance in bits between the eviction probability distributions $p_e(a)$ 
and $p_u(a)$ using the Kullback-Leibler (KL) divergence to measure, 
\begin{equation}
  D_{\mathit{KL}}(p_e||p_u) =  \sum_{a\in [0,...,N-1]} p_e(a) \log_2 \frac{p_e(a)}{p_u(a)}.   
\end{equation}
Note that the KL divergence does not fulfill the requirements of a metric and is asymmetric. Nevertheless, $D_{KL}(p_e||p_u)$ describes the relative entropy of $p_e(a)$ with respect to $p_u(a)$ and is a measure of the information lost if $p_u(a)$ was used to approximate $p_e(a)$. Mapped to cache side channels, the KL divergence thus nicely characterizes the leakage of a cache design with an eviction probability distribution $p_e(a)$ relative to the distribution $p_u(a)$ in a fully associative cache design.

\parhead{Sampling $p_e(a)$.}
To sample $p_e(a)$, we simply count the number of evictions for the attacker's cache lines when the victim repeatedly accesses a fixed, randomly chosen address. More specifically, we first fill the cache by randomly accessing cache line addresses from the memory chunk corresponding to the attacker's security domain.
To keep track of self-evictions and hence the attacker's lines that are actually cached, we utilize our cache model's capability to return which cache line is evicted with each access, as described in \cref{sec:cache-model}. We note that this is an over-approximation of the attacker's capabilities, as on real systems this translates to an attacker who can perfectly monitor cache evictions and accurately determine address collisions in the cache.
Once the cache is entirely filled with the attacker's data, we access a fixed secret address from the victim's security domain, forcing an eviction of one of the attacker's addresses. We then increment 
 the eviction counter for the attacker address that is being reported as evicted from the cache. 
We repeat this sampling step multiple times and finally divide the per-address eviction counts by the total number of observed evictions, thereby obtaining $p_e(a)$. The repeated sampling procedure reduces the error of the sampled eviction probabilities proportional to $\sqrt(r)$, where $r$ is the number of samples collected.

\begin{figure}
 \begin{center}
 \includegraphics[width=\columnwidth]{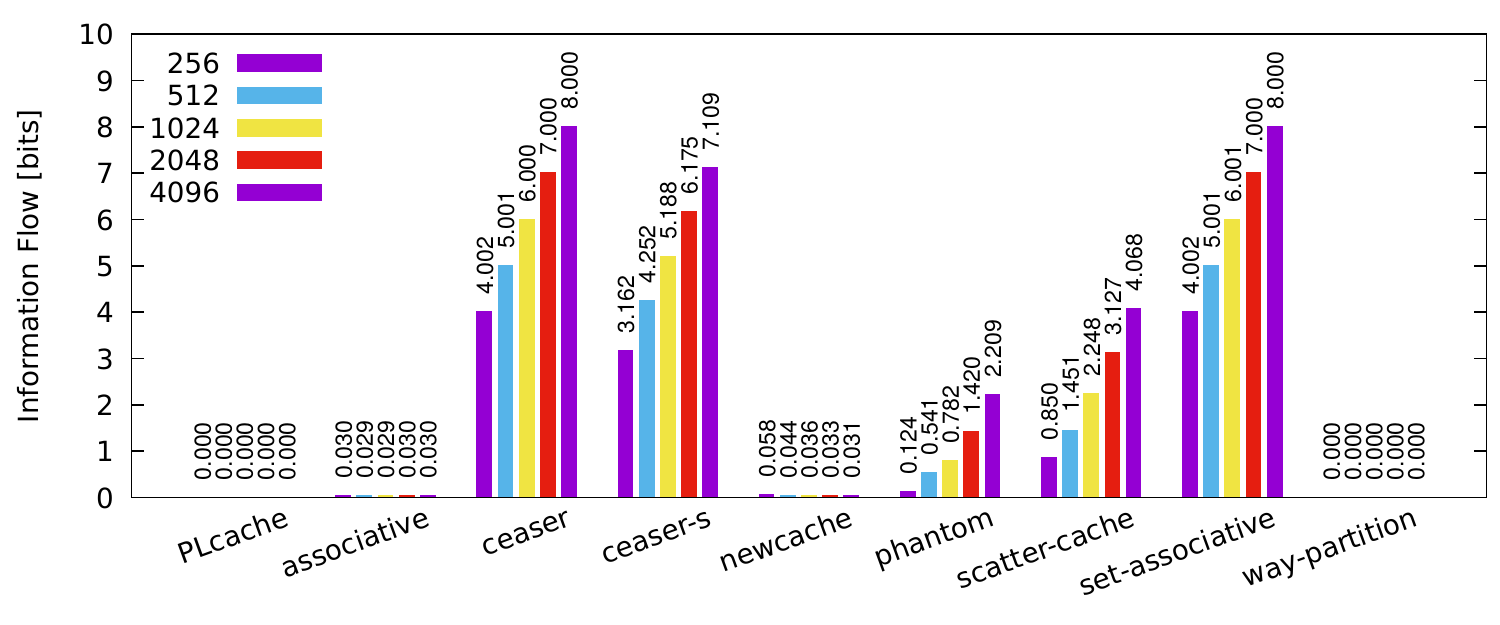}
 \end{center}
 \caption{REE for different cache designs with random replacement. All but NewCache and the fully associative cache use 16 ways.}
 \label{fig:entropy_overview_rand}
\end{figure} 

\parhead{Evaluation Results.}
\autoref{fig:entropy_overview_rand} depicts the information leakage in the analyzed cache designs for various cache sizes and using random replacement.
While the partitioned cache designs exhibit zero leakage, the leakages for CEASER and set-associative caches is the number of sets, i.e., $\log_2(\#{sets})$ bits, thereby confirming the validity of our results. Next, we attribute the slightly above-zero leakages in NewCache and the fully associative cache to statistical noise.
Note that CEASER-S and ScatterCache (with 2 and 16 partitions, respectively) show considerably lower leakage than standard set-associative caches. 
Moreover, as PhantomCache is looking up 8 sets, i.e., 128 lines, in parallel, PhantomCache stands out with significantly lower leakage per access than other designs, but also hurts chip area and power consumption.

\begin{figure}
 \begin{center}
 \includegraphics[width=\columnwidth]{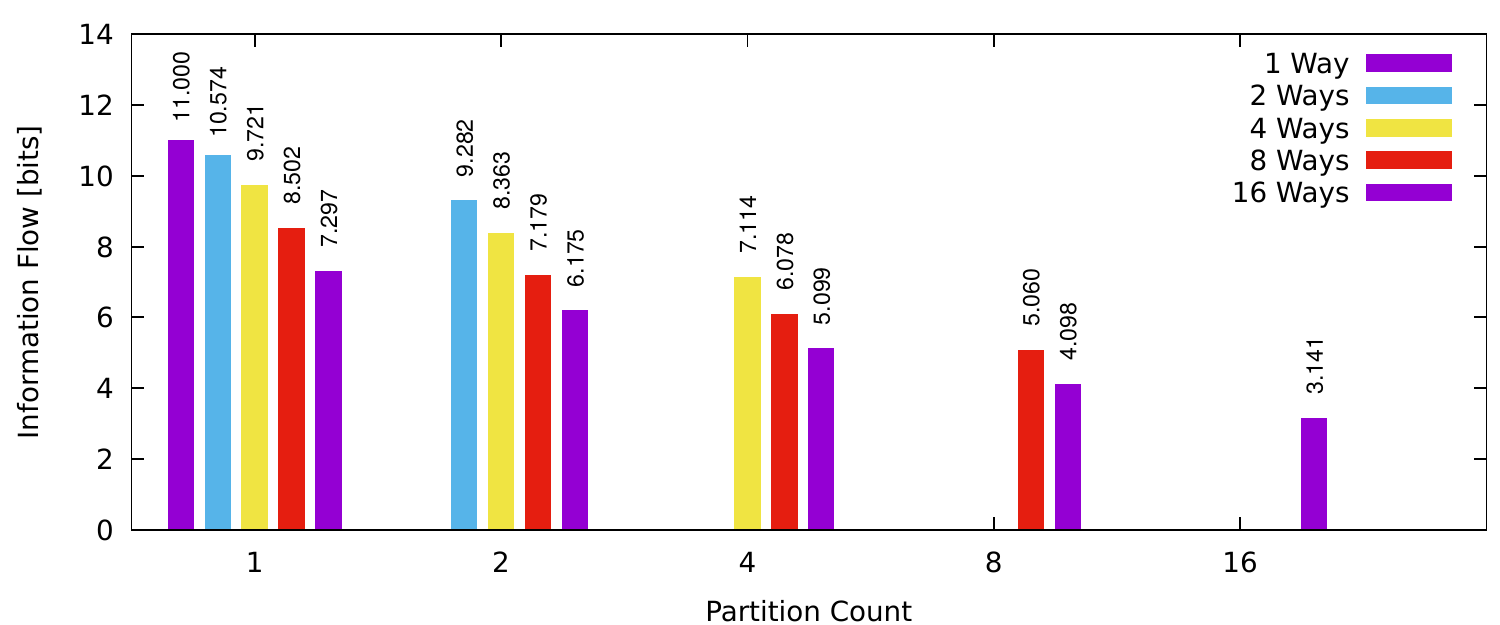}
 \end{center}
 \caption{REE for CEASER-S with 2048 lines depending on ways and partitions.}
 \label{fig:entropy_ceasers_2048}
\end{figure} 

\autoref{fig:entropy_ceasers_2048} analyzes the leakage in skewed caches 
like CEASER-S depending on way and partition count. \autoref{fig:entropy_ceasers_2048} clearly shows that increasing the number of ways and partitions effectively reduces leakage, with the difference between the best and worst configuration being 8\,bits per access.

\parhead{Supporting More General Cache Designs.}
We note that our Relative Eviction Entropy method can be computed in linear time, allowing us to evaluate different cache designs within minutes. However, we do assume some properties of the replacement policy of the cache being tested, namely that every line in the considered cache design exhibits the same leakage behavior,  which in turn is independent from the specific address accessed by the victim.  We  rely on this assumption in  our procedure for sampling $p_e(a)$, evaluating the eviction distribution using only a single fixed address accessed by the victim.
We argue that this assumption is natural and holds for most cache designs, including all the caches considered in this paper, as typical replacement policies do not differentiate between cache line addresses. While a single access does not reflect practical attack scenarios, it gives strong insight into the theoretical leakage caused by the caches's structural mapping of addresses to cache lines. To better understand the practical exploitability of this leakage, we conduct application-specific tests using cryptographic routines later in \autoref{sec:eviction-set-attacks}.
 However, note that the REE metric can be easily adapted to other cases as well, by simply testing multiple victim addresses and reporting the range of the occurring leakage as a function of victim's address.

\subsection{Eviction-Set Creation}\label{sec:evict-create}
To perform contention-based cache attacks, attackers first
construct suitable eviction sets, i.e., minimal sets of
addresses in their own address space that collide with the victim's accesses of
interest.  
Due to its perceived importance, multiple cache designs aim at randomizing
the cache to prevent efficient eviction-set creation and thus
contention-based attacks.

\parhead{Constructing Eviction Sets on Randomized Caches.} 
Previous works proposed a range of methods for finding
eviction sets in randomized caches. 
Taking a top-down approach, the Single Holdout Method
(SHM) and the Group Elimination Method (GEM)~\cite{Qureshi19,VilaKM19} both
start from a large set of attacker addresses that evicts a certain victim
address and then shrink this conflict set to a minimal eviction set by
trying to remove (groups of) addresses while continuously verifying that 
the cache conflict remains. Taking a bottom-up approach, 
the Prime+Prune+Probe (PPP) method~\cite{PurnalGGV21,Purnal19}
pre-fills the cache with a set of candidate addresses, and subsequently
triggers the victim access of interest. PPP then tests for cache
misses in its candidate set, thereby locating conflicting addresses. Note 
that all of these approaches allow for optimizations specific to the 
cache replacement strategy in use.

\parhead{Evaluating Difficulty of Eviction Set Construction.}
As protecting against eviction set construction is a major design goal for
randomized caches, \fw allows to evaluate the effectiveness of
SHM, GEM, and PPP on a candidate cache design. %
In particular, \fw allows us to quantify the number of memory accesses required by
an attacker, the number of conflicting addresses found, and
the success rate of using the found addresses for evicting the victim address.
These figures eventually allow to configure cache re-keying intervals, e.g.,  for 
CEASER and CEASER-S. To set a level playing field and support an equal comparison across cache designs,
we use the same implementations of eviction-set construction techniques for all 
evaluated cache designs.
We intentionally avoided cache-specific optimizations, opting for comparable results rather than for optimal strategies.
Specifcally, all of our implementations iterate until they find (or shrink a
conflict set to) the minimum number of addresses required for an eviction set,
or until a predefined maximum iteration count is reached. The latter is a necessity to
perform bulk testing as some algorithms do not terminate for
every cache design.

\parhead{Measurement Setup.}
To measure the success rate, we set up a clean cache environment 1000 times and
count the number of successful evictions of the cached victim address given the
found eviction set. We extracted the cache hit/miss statistics to evaluate the
number of attacker accesses needed for eviction-set creation. We determine the number 
of true conflicts in the eviction set by testing every found address for a 
collision with the victim address in the cache. 
While this is not directly possible on real systems, \fw provides this
feature to assess how well each algorithm works for every cache
design.

In our experiments, we used random replacement, 2048 lines and 16 ways where applicable, i.e.,
except for NewCache and the fully associative cache, which only have one set. We operated CEASER-S with 2 partitions, 
NewCache with $k=2$, and PhantomCache with 8 parallel set lookups. We set up the
algorithms to look for as many addresses as there are cache ways. For PhantomCache, however, we require 8x the number of ways, because it can place lines
in 8 different sets.

\begin{figure}[htb]
	\begin{center}
		\includegraphics[width=\columnwidth]{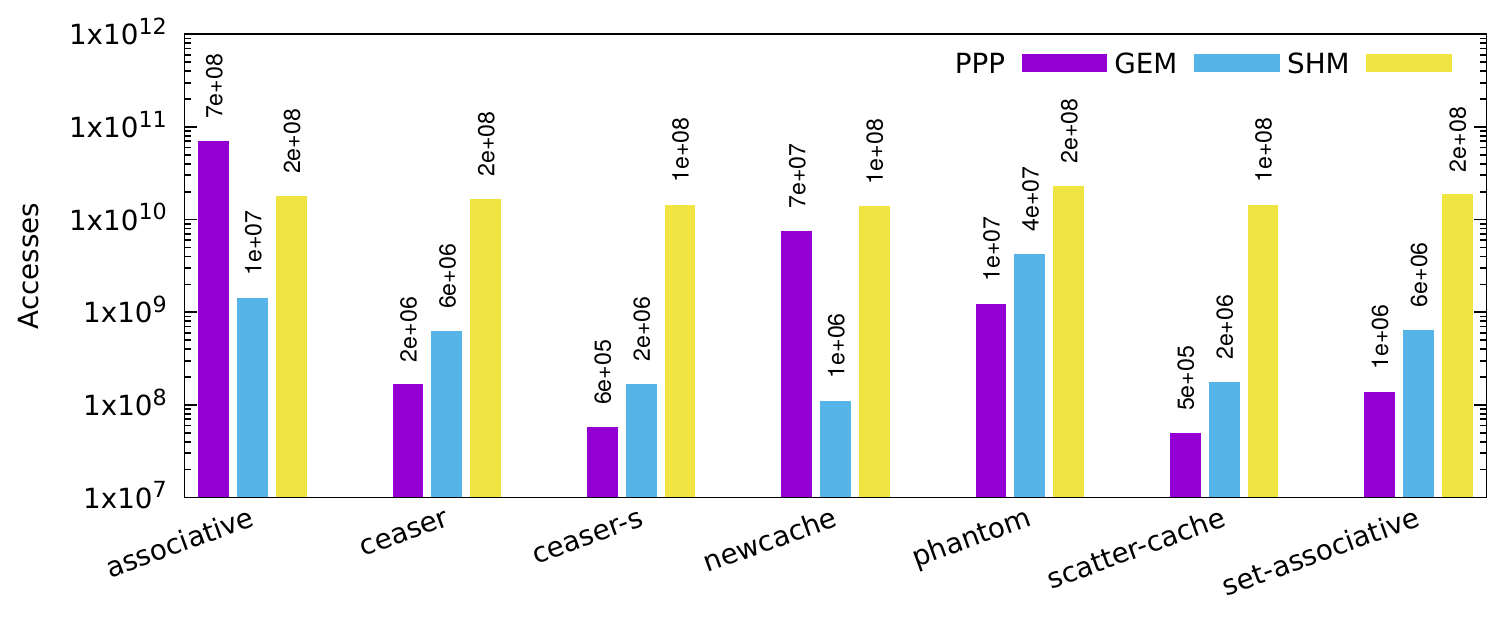}
	\end{center}
	\caption{Number of memory accesses required by eviction-set building techniques for different 2048-line caches.}
	\label{fig:evset_build_overview_accesses}
\end{figure}

\parhead{Evaluating the Number of Memory Accesses for Eviction Set Construction.}
\autoref{fig:evset_build_overview_accesses} shows the number of memory operations 
done by SHM, GEM, and PPP for different cache designs.
As L1 cache accesses take about five CPU cycles, these results give an 
indication about the execution time of each technique when used against a specific cache design.

\begin{figure}[htb]
	\begin{center}
		\includegraphics[width=\columnwidth]{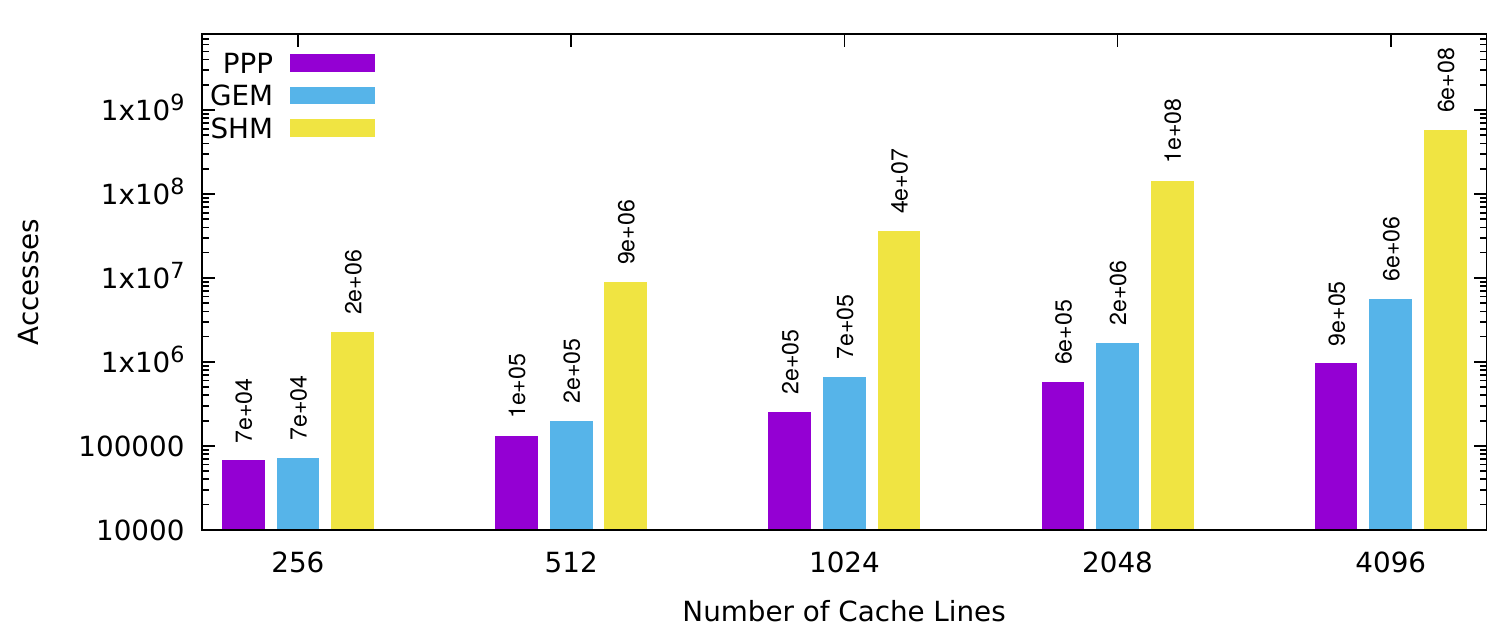}
	\end{center}
	\vspace{-1em}
	\caption{Number of memory accesses required by eviction-set building techniques for CEASER-S depending on cache size.}
	\label{fig:evset_build_ceaser_s_accesses}
\end{figure}

As the figure shows, the number of memory accesses for SHM is the highest, and in
the same order of magnitude for all designs. In contrast, the complexity of PPP
scales with the eviction set size, e.g., PPP is two orders of magnitude faster 
for ScatterCache than for NewCache. We also observe that 
PPP tends to be more efficient for skewed caches, as it is~3x faster for
CEASER-S than for CEASER. 
The performance of GEM is mostly in between PPP and SHM, but tends to be faster
than PPP in the case of large eviction sets (e.g., for NewCache). Finally,
\autoref{fig:evset_build_ceaser_s_accesses} gives additional performance
figures for CEASER-S and demonstrates the linear increase in complexity with
the number of cache lines.

\begin{figure}[htb]
	\begin{center}
		\includegraphics[width=\columnwidth]{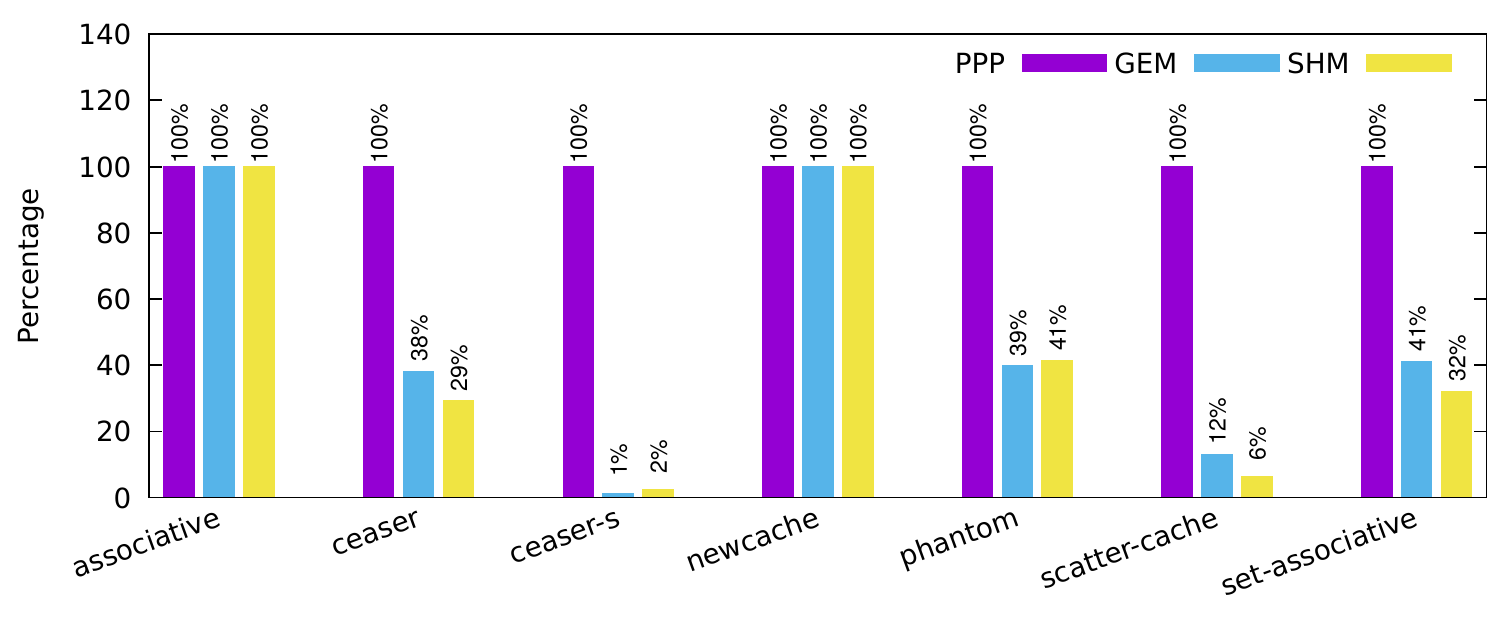}
	\end{center}
	\vspace{-1em}
	\caption{Percentage of addresses in the constructed eviction sets that conflict with the victim's address, using different eviction-set construction techniques and 2048-line caches.}
	\label{fig:evset_build_overview_tpr}
\end{figure}

\parhead{Evaluating Eviction Coverage.}
Different eviction set construction techniques can also produce eviction sets of different quality.
\autoref{fig:evset_build_overview_tpr} thus shows the number of addresses in
the found eviction sets that truly conflict with the victim address: PPP works 
best for all of the tested cache designs, producing eviction sets
where all of its addresses truly conflict with the victim address.  
In contrast, SHM and GEM are less reliable, 
producing eviction sets where many of the addresses do not conflict with the victim address. 
The main reason for this is that SHM and GEM are highly susceptible to noise, 
which stems from both random replacement and cache skewing.

\begin{figure}[htb]
 	\begin{center}
 		\includegraphics[width=\columnwidth]{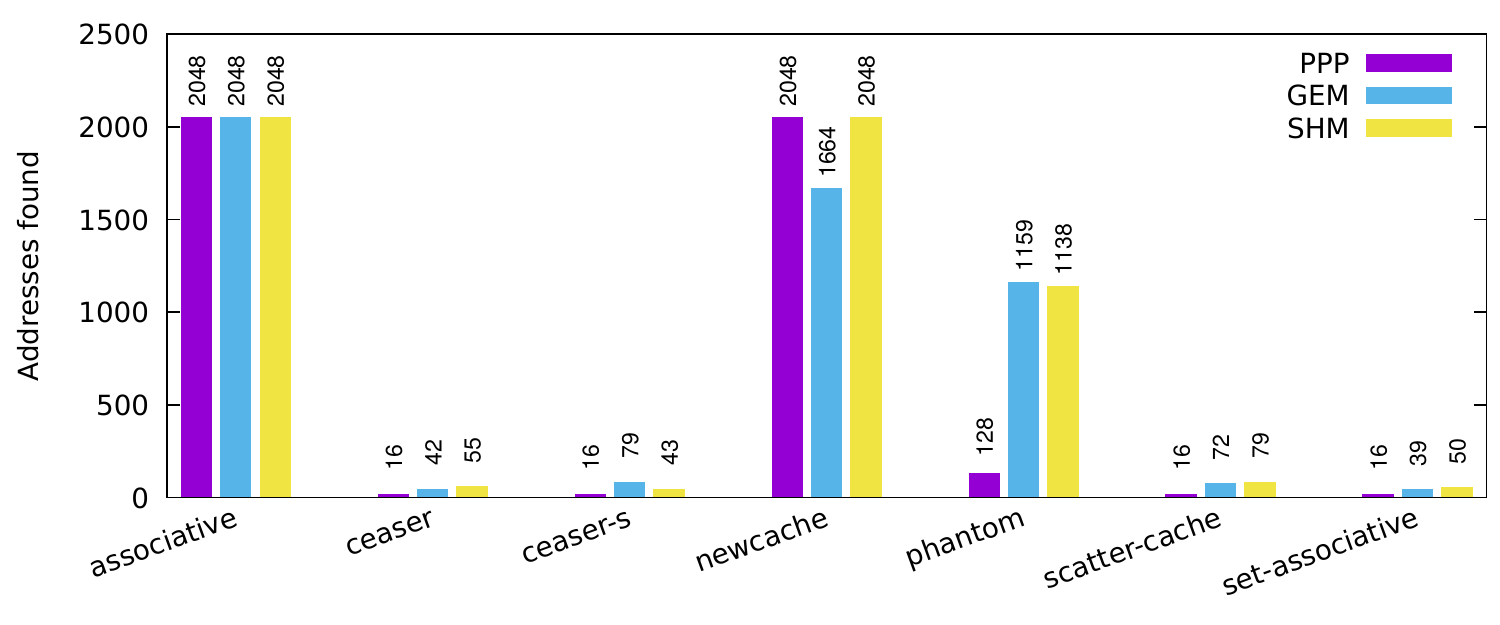}
 	\end{center}
	\vspace{-1em}
 	\caption{Eviction set sizes found by eviction-set building techniques for different 2048-line caches.}
 	\label{fig:evset_build_overview_addrFound}
 \end{figure}

To verify this, \autoref{fig:evset_build_overview_addrFound} shows the constructed eviction 
sets' sizes for SHM, GEM and PPP. Except 
for NewCache and fully associative caches, both SHM and GEM stop shrinking the
conflict set before it becomes minimal, which results in eviction sets where
many addresses do not conflict with the victim address. This effect is
particularly strong for the skewed cache designs CEASER-S and ScatterCache.
Moreover, SHM and GEM also fail on PhantomCache, where both
algorithms terminate with 10x as many addresses as needed.
Finally, for NewCache and fully associative caches every address is equally suitable for an
eviction set, which automatically results in 100\% of the addresses conflicting
with the victim address.

\begin{figure}[htb]
	\begin{center}
		\includegraphics[width=\columnwidth]{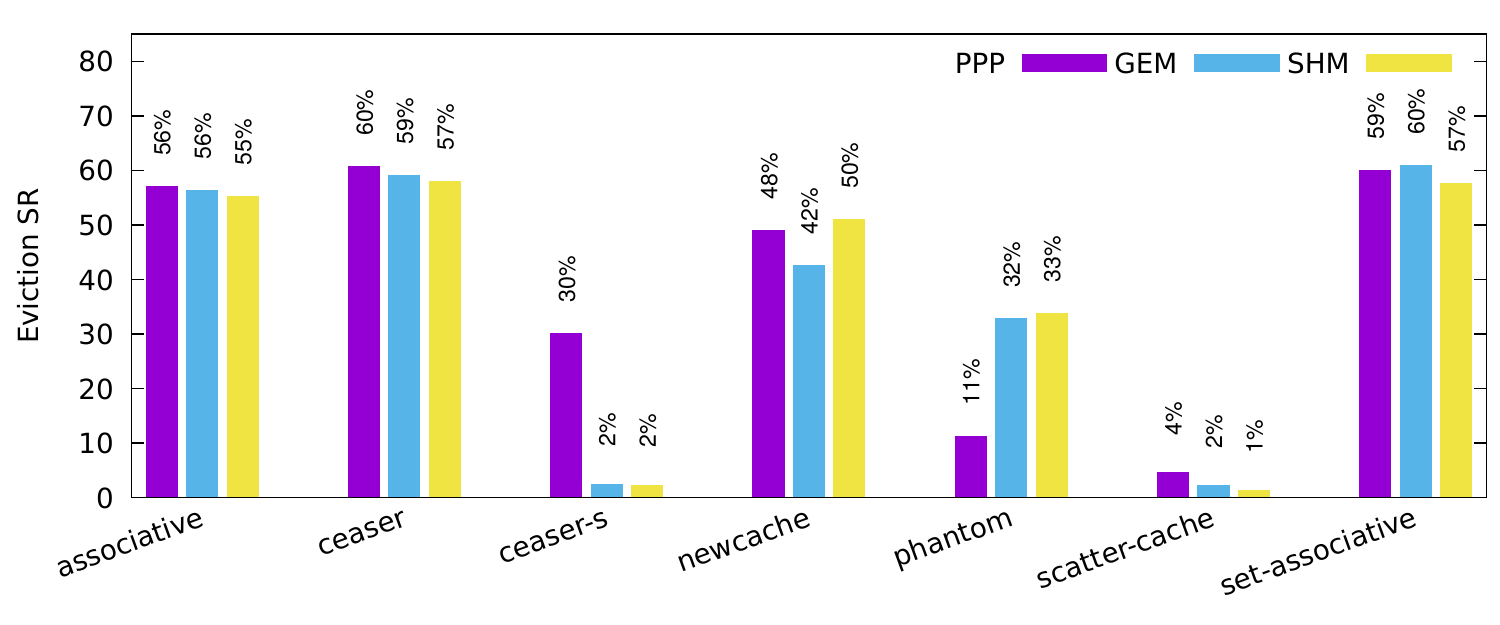}
	\end{center}
	\vspace{-1em}
	\caption{Eviction success rate for the eviction sets found for different 2048-line caches.}
	\label{fig:evset_build_overview_successrate}
\end{figure}

 \parhead{Evaluating Eviction Success Rate.}
We also evaluate the constructed eviction sets for their ability to effectively
evict the victim address of interest. 
As \autoref{fig:evset_build_overview_successrate} shows, the eviction sets found by all
three eviction set construction techniques perform equally well for CEASER,
NewCache, set- and fully associative caches. For CEASER-S and ScatterCache, PPP
yields better eviction success rates than SHM and GEM, because PPP is generally
more accurate (cf. \autoref{fig:evset_build_overview_tpr}). For PhantomCache,
however, GEM and SHM yielded better eviction rates as the found eviction set
makes up roughly 50\% of the cache. As skewed caches exhibit a significantly
smaller probability of successful eviction (e.g., 2-4\% for ScatterCache), 
eviction sets might be chosen larger to obtain high eviction probabilities and 
and \pp observability. 

\begin{figure}[htb]
 \begin{center}
 \includegraphics[width=\columnwidth]{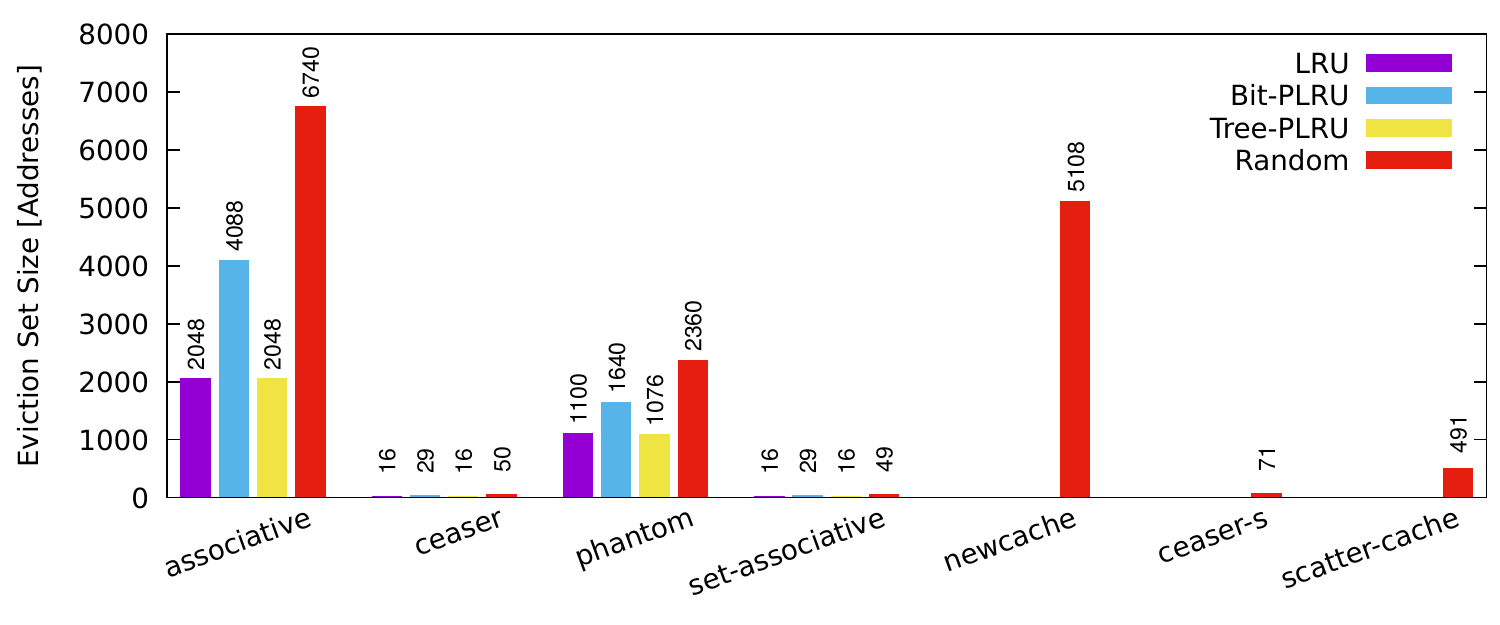}
 \end{center}
	\vspace{-1em}
 \caption{Eviction set size for 2048-line caches and $90\%$ eviction probability.}
 \label{fig:evset_build_overview_evset_size}
\end{figure}

\parhead{Obtaining a Specific Eviction Probability.}
To learn how many addresses would be needed to yield a certain
eviction probability $\alpha$, we start with an empty eviction set and
successively add conflicting addresses until the eviction probability reaches
$\alpha$. \autoref{fig:evset_build_overview_evset_size} presents the results of
this routine for $\alpha=90\%$, across different caches and replacement
policies. It shows that LRU and Tree-PLRU allow for smaller eviction sets than
Bit-PLRU and random replacement. In addition, skewing significantly increases
the number of conflicting addresses needed, e.g., ScatterCache requires 10x
more addresses than CEASER with equal sets and ways.

\subsection{Eviction-Set Attack}\label{sec:eviction-set-attacks}
In this section we shift the focus from observing specific aspects of the cache
to measuring the security offered by
emerging cache designs when performing attacks on cryptographic
implementations. To that aim, we simulate victims that use a cryptographic
algorithm while the attacker tries to learn enough information
to distinguish between two keys used by the victim. We use two
cryptographic algorithms, each representing a different type of cache attack.

\parhead{The AES Victim.}
Our AES victim is based on code from OpenSSL, which uses a set of tables,
called {T-tables}, implemented as arrays.  
The attack focuses on the first four accesses made to the
first T-table during the encryption.  The two keys are selected
such that, when encrypting some \emph{vulnerable} plaintexts with the first
key, all of these accesses fall in the first cache line of the T-table.
Conversely, when encrypting vulnerable plaintexts with the second key, each of
the four accesses falls in a different cache line.  Finally, to further
facilitate the attack, we allow the attacker to choose as many  random
vulnerable plaintexts as required for the attack. 
In a more realistic scenario, the attacker can guess the characteristics of the vulnerable plaintexts.
Thus, allowing to select vulnerable plaintexts represents a constant factor improvement in attack complexity.

\parhead{Modular Exponentiation Victim}
Our second victim implements modular exponentiation,  a core operation in
multiple public-key schemes, e.g., RSA.  Our modular exponentiation victim
gets a 2048-bit base $b$, a 2048-bit modulus $m$, and a 32-bit exponent $e$.
The victim then uses the square-and-multiply algorithm~\cite{GORDON98}  to calculate
$b^e \bmod m$.  The square-and-multiply algorithm maintains an accumulator $a$
that is initialized to 1.  For each bit of the exponent $e$, the algorithm squares $a$, and if the
bit is set the algorithm also multiplies $a$ by $b$, reducing $a$ modulo $m$ as necessary.
Thus, the multiplication code is only executed when the exponent bit is 1.

The keys are selected so that the value of a bit at a specific index (7 in our tests)
of the exponent is 0 in the first key and 1 in the second. 
The other bits of each exponent are randomly chosen.
We simulate an attacker that runs concurrently with the
victim.  The attacker can manipulate the cache whenever the victim finishes
processing an exponent bit.

\parhead{Attacker Setup }
In the attack setup phase, the attacker is provided with an eviction set that evicts
a monitored victim cache line with a probability 90\%.
We construct this eviction set by successively adding conflicting addresses to 
an initially empty set as outlined in \autoref{sec:evict-create}. 
See there for a complexity analysis of eviction-set construction.

\parhead{Attacker Procedure.}
The attack proceeds as a sequence of rounds. In each round, the attacker 
asks
the victim to encrypt a plaintext with the two selected keys, randomizing the
order of using the keys in each round 
to avoid cache effects that depend on the order of the use of keys.
Before each encryption, the attacker accesses the eviction set three times to prime the cache.
After each encryption, the attacker accesses the eviction set, counting the number
of cache misses during these accesses.
Finally, the attacker calculates the average number of cache misses for each
key, and stops when achieving a 95\% confidence that the averages differ, or
when hitting a predefined number of rounds.  (1,000 for the modular
exponentiation and 100,000 for AES.)
To overcome the case where the eviction set and the victim all fit in the cache, the attacker accesses some arbitrary memory when no cache evictions are observed during a round.

\parhead{Selecting Cache Designs for Evaluation.}
We perform the attack on a sample of the cache designs considered in this paper. 
First, we do not test partitioned caches, 
because these do not leak information as there is no resource contention 
between the attacker and the victim. 
Secondly, to ensure that results are comparable,
we limit our experiments to a cache size of 256 lines. 
Where applicable, we vary the associativity, testing all powers of two 
between 1 and 16. 
For each configuration, we run the attack 1,000 times and report the median 
of the number of encryptions required for distinguishing the keys. 
We use the median rather than the mean because in some cases the distribution has a long tail, skewing the mean towards a small number of cases where many encryptions are required.

\begin{figure}[htb]
 \begin{center}
 \includegraphics[width=\columnwidth]{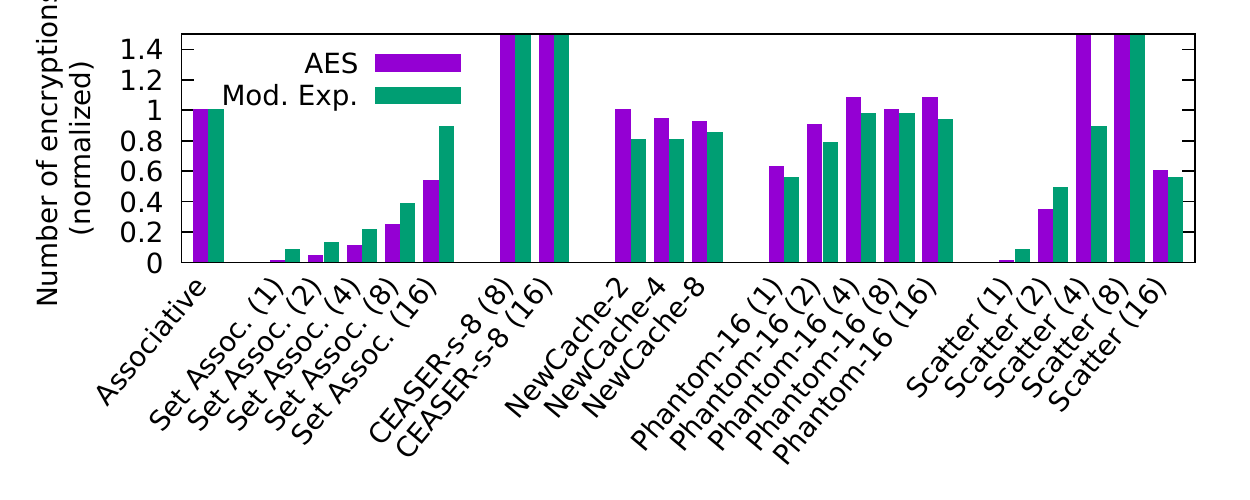}
 \end{center}
  \vspace{-1em}
  \caption{Eviction-set attack: Number of encryptions required to break AES and modular exponentiation
  with random replacement.
  CESER, CEASER-s-1, and Phantom-1, which show behavior similar to set associative caches, have been omitted from the figure.
  (Normalized to a random-replacement associative cache.)}
 \label{fig:eviction-rand-enc}
\end{figure}

\parhead{Observing Key Leakage.}
\autoref{fig:eviction-rand-enc} shows the
median number of encryptions required 
for the attacks when using a random replacement strategy.
We normalize the results to the number of encryptions required for
the fully associative cache. (10,590 and 94 for AES and modular exponentiation,
respectively.)
For brevity, we also omit the results of 
CEASER, CEASER-S with one partition, and PhantomCache with one set lookup, 
all of which  do not seem to offer any advantage over a set-associative cache with the same associativity. 

The figure shows that all NewCache variants and PhantomCache with 16 set lookup are mostly equivalent to the fully associative cache.
CEASER-S with 8 partitions provides a stronger protection.
In particular, the majority of the AES attacks on this cache design with 8 ways and of the
modular exponentiation attacks with 16 ways were not successful.

The results with ScatterCache are a mixed bag.  When the associativity
is four or eight, the design provides a good protection, equivalent or surpassing
the fully associative cache.
(In particular, the AES attack fails in most cases on an 8-way cache.)
However, the protection is lower for the other cases.

\subsection{Cache-Occupancy Attack}\label{sec:occupancy}
We now turn our attention to an emerging cache attack strategy
that ignores spatial information and instead only utilizes the
victim's overall cache usage~\cite{ShustermanKHMMO19,MauriceNHF15,ShustermanAOGOY21}.
To measure resistance against
so called \emph{cache-occupancy attacks}, we use the same cryptographic victims as
in \cref{sec:eviction-set-attacks}.
The attacker is still tasked with distinguishing between two keys,
but instead of using an eviction set targeting a specific cache line,
the attacker uses a cache-size buffer and counts the number of cache misses
when scanning the buffer.
(A different sized buffer may also work~\cite{ShustermanACHKL21}, but this requires further investigation.)
Most other aspects of the attack are the same as in our eviction-set attack.
We do not, however, handle failed eviction because using a cache-size buffer
guarantees contention on the cache.

\begin{figure}[htb]
 \begin{center}
 \includegraphics[width=\columnwidth]{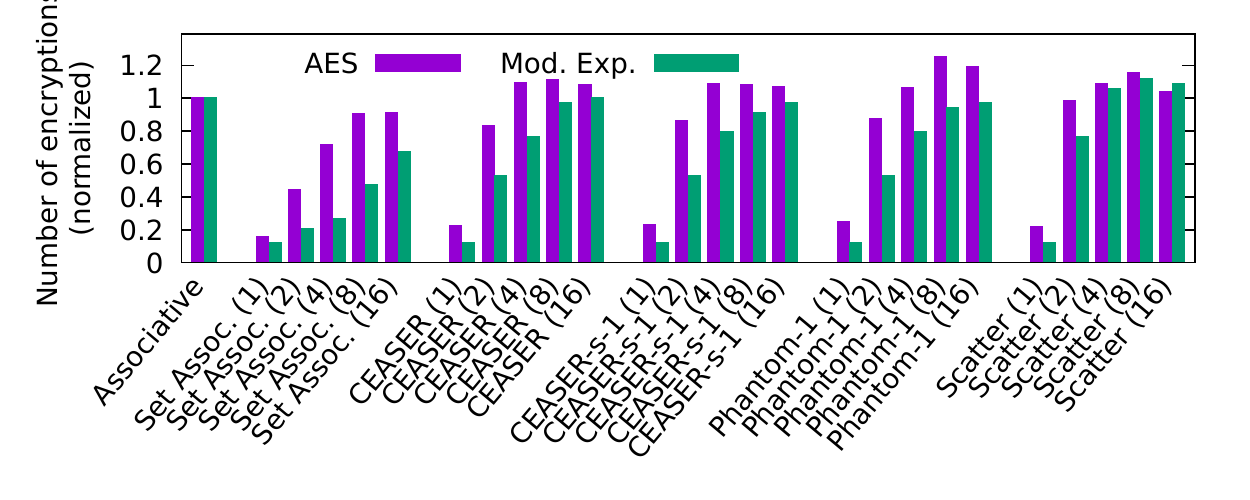}
 \end{center}
	\vspace{-1em}
  \caption{Occupancy attack: Number of encryptions required to break AES and modular exponentiation
  with random replacement. CEASER-S-8, NewCache, and Phantom-16, which show behavior similar to fully associative caches, have been omitted from the figure.
  (Normalized to a random-replacement associative cache.)}
 \label{fig:occupancy-rand}
\end{figure} 

\parhead{Observing Key Leakage.}
\autoref{fig:occupancy-rand} shows the
median number of encryptions required for
the cache occupancy attacks when using a random replacement strategy.
As in \cref{fig:eviction-rand-enc}, we normalize the results to the 
number of encryptions required for
the fully associative cache. 
Similar to the eviction-set attack, NewCache, CEASER-S with 8 partitions, and PhantomCache with 16 set lookup achieve a protection similar to that of fully associative cache. 
(We have omitted these three from the figure for brevity.)
Most other configurations achieve a protection level which is significantly better than set-associative caches, in particular for the attack on AES.

Due to normalization, \cref{fig:eviction-rand-enc,fig:occupancy-rand} do not show that occupancy attacks on the fully associative cache 
require significantly less encryptions than eviction-set attacks.
(5664 and 68 for AES and modular exponentiation, compared to 10590 and 94.)
The cause is that the eviction set algorithm targets 90\% eviction rate, which for fully associative caches leads to eviction sets 
that are larger than the cache-sized buffer used in the occupancy attack and thus more self evictions.

\begin{figure}[htb]
 \begin{center}
 \includegraphics[width=\columnwidth]{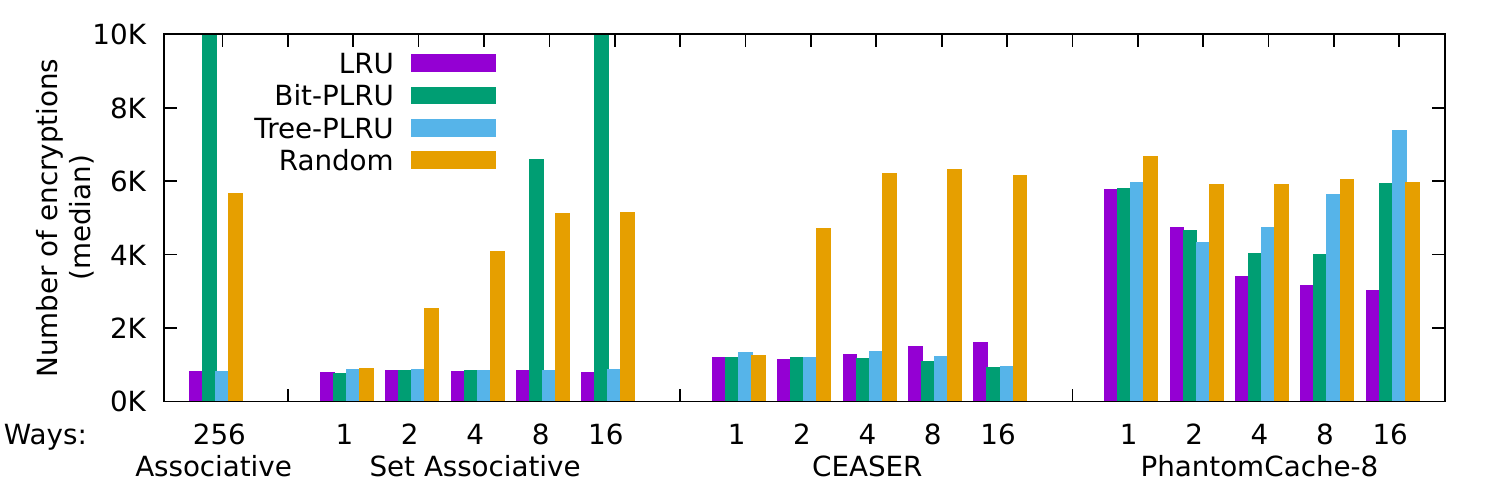}
 \end{center}
	\vspace{-1em}
  \caption{Median number of encryptions required to break 
  AES with different cache designs and replacement algorithms.
  Fully associative and 16-way set associative caches are not fully
  represented, requiring 16,984 and 22,116 encryptions for Bit-PLRU, respectively.
  }
 \label{fig:occupancy-aes}
\end{figure} 

\parhead{Comparing Different Replacement Algorithms.}
\cref{fig:occupancy-aes} shows the effect of changing the replacement policy
on the attack complexity.
As the figure demonstrates, in most cases, caches with a random replacement policy
offer significantly better protection than those with deterministic replacement.

For deterministic replacement policies, we observe that CEASER only provides marginal benefit over set-associative caches, whereas PhantomCache provides a significantly better protection than other cache designs.
We believe that the reason is that PhantomCache is inherently non-deterministic,
hence, even with deterministic replacement algorithms, PhantomCache can reduce the correlation between the victim's access and the attacker's observation.

Attacks on deterministic cache designs that use bit-based pseudo-LRU replacement exhibit an anomaly that increases the number of encryptions required for statistical confidence.
The cause is that the algorithm experiences some rare cases where a single cache miss causes cascading evictions of the eviction set. 
These rare cases increase the variance of the number of evictions observed, and with it the number of samples required.
Modifying the attack to ignore outliers will eliminate these rare cases and significantly improve the attack.
Hence, the results do not indicate that bit-based pseudo-LRU is more secure than random replacement.

\subsection{Optimal Eviction-Set Size}
In \cref{sec:evict-create} we evaluate eviction sets based on the probability of evicting a victim cache line from the cache.
However, as discussed in \cref{sec:occupancy}, 
larger eviction sets can result in lower attack efficiency.
The main reason for the observed reduction is self evictions.
Specifically, increasing the size of the eviction set increases the probability of cache conflicts between elements of the eviction set.
These self evictions introduce measurement noise that increases the variance in the measurements and consequently the number of samples the attacker needs to observe to distinguish the keys.
As a secondary effect, larger eviction sets require more memory accesses for both the prime and the probe steps of the attack, reducing attack efficiency.

\begin{figure}
	\begin{center}
		\includegraphics[width=\columnwidth]{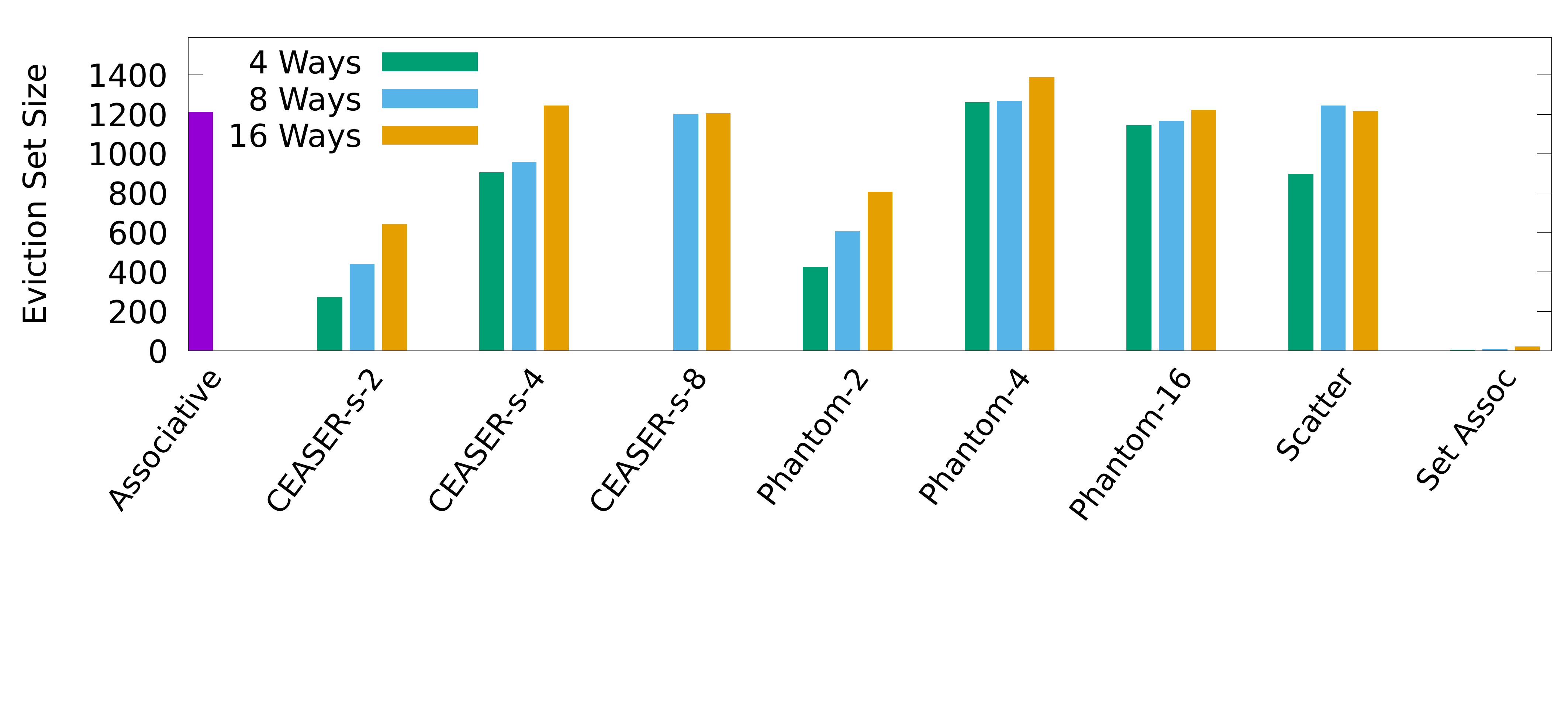}
	\end{center}
	\vspace{-1em}
	\caption{Optimal eviction set sizes for different caches with 1024 lines}
	\label{fig:cl1024_w16_optimal_eviction_set_sizes}
\end{figure} 

As a final example of the flexibility of \fw, we now use it to find the eviction-set size that allows for the most efficient attack.
Specifically, we experiment with various cache designs, all with size 1024,
our AES victim, and our eviction-set attacker.
We vary the eviction-set size between 1 and 2048, and measure the median number of encryptions required for distinguishing the keys.
\cref{fig:cl1024_w16_optimal_eviction_set_sizes} reports the eviction-set size that allows the attack with the minimal median.
As we can see, a lower associativity allows for smaller eviction-set sizes.
However, when the associativity grows to 16, in most cache designs the best eviction-set size is similar to that of a fully associative cache, indicating that occupancy-based attacks are as effective as eviction-set attacks.

\section{Threats to Validity and Limitations}\label{s:validity}
At the moment, \fw does not support the evaluation of cache hierarchies.
Consequently, designs that rely on the hierarchy for defense are
outside the scope for this work. Moreover, evaluations using 
\fw currently assume a noise-free scenario, which provides a 
conservative security estimate as the absence of noise is the best case for attackers. 
However, practical cache attacks also face systematic and random noise
stemming from other system activity. 
To assess the impact of noise and cache hierarchies to security, we aim 
to add such models to \fw in the future. 

For our cryptographic attack evaluations, \fw models a strong, synchronized
attacker and an artifical victim that computes (and leaks) upon the attacker's
request. As for noise, this is a very strong attack model that allows
to obtain a lower bound for security. While a simple model like \fw cannot 
capture all complexities involved in real-world attacks, we consider modeling
more realistic scenarios as future work. 

Another relevant aspect of a secure cache is its performance. At this point,
\fw does not support evaluation of cache performance, but its 
mechanisms to collect data about memory accesses, cache hits and misses 
allows in the future to easily extend \fw to measure, e.g., the cache hit rate, 
based on memory traces of relevant workloads.

\section{Related Works}\label{s:related}
Past work on evaluating the security of caches against side channel attacks mainly focused on three aspects: 1) formal model of cache and theoretical analysis of information leakage, 2) metrics for empirical quantification of information leakage, 3) modeling of cache side channel attacks.

\parhead{Formal Cache Model and Theoretical Analysis.} This line of research \cite{KopfMO12, DoychevFKMR13} tries to formally model the state change of the cache and extend the program execution semantics to include cache state changes by leveraging prior work on formal analysis of cache miss rates. Eventually they can estimate the number of reachable cache states and give an upper bound on the leakage in terms of channel capacity, for a given program under analysis. Similarly, \cite{Ghasempouri2020} models caches and cache attacks as automata to verify cache security using model checking.
Due to the restrictions of formal methods, these works are limited to simple cache models (e.g. set-associative cache with LRU replacement) and can only give a very loose upper bound of leakage. Hence, they are not suitable for comparing the security of various complex secure cache designs. 
In contrast, \fw empirically evaluates a number of metrics to quantify side-channel leakage in software cache models and evaluates the exploitabiliy of cache leakage for programs such as cryptographic algorithms.

\parhead{Metrics for Empirical Quantification of Information Leakage.}  
Another line of research introduces metrics to empirically evaluate the security of cache designs and implementations, such as by using mutual information and min-leakage~\cite{CockGMH14}, by using a linear correlation coefficient between oracle traces and the attacker's observations \cite{DemmeMWS12, DemmeMWS13, ZhangLCL13, DemmeS14}, by measuring the accuracy of deep learning models trained to learn the relationship between victim accesses and the attacker's cache observations~\cite{ZhangZL18}, or by modeling and statistically analyzing cache side channels using communication theory~\cite{BourgeatDYTEY20}.
\fw as well tries to empirically characterize the leakage of cache designs. However, as we point out, a single metric is insufficient to entirely capture cache security. Moreover, none of these works looks at cache occupancy channels or tries to assess security by using well-studied cryptographic targets. 

\parhead{Modeling of Cache Side Channel Attacks.}
Some works tried to model caches and cache attacks such as to detect and quantify cache leakage. For instance, \mbox{\citet{ZhangL14}} model the cache as a finite state machine to identify interference and determine the mutual information. \citet{HeL17} model cache attacks as a Probabilistic Information Flow Graph (PIFG) to derive for each cache and attack an overall probability of success. \citet{WangZWM19} derive a risk score from modeling attacks using Petri nets and calculating the success probabilities of concrete attacks. \citet{DengXS19,Deng0S20,DengMXKS21} model cache attacks as a series of three consecutive read/invalidation steps, identify vulnerable three-step patterns using a simulator, and use the model for evaluating the security of the caches in multiple Arm devices. In addition, their work introduces a Cache Timing Vulnerability Score (CTVS) from running vulnerable patterns on real machines. While these prior works greatly improve the understanding of cache attacks, many are based on simple cache models. \fw thus takes another step forward and automatically evaluates arbitrary software models of cache designs w.r.t.\ to a number of different metrics and attack complexity to provide a comprehensive security report.

\section{Conclusion}\label{s:conclusion}
This work fills a gap in the practice of evaluating cache designs for security.
It presents \fw, a flexible framework that supports multiple metrics.
We experiment with three different, albeit related, metrics, 
which we use to evaluate and compare 
multiple secure cache designs.

We observe that all of the non-partitioned caches leak information
and note that the leak is sufficient to implement cryptographic attacks.
Moreover, as predicted, we show that a single metric may fail to capture all of the intricacies.
For example, the REE of CEASER-S indicates less leakage when the number of ways or partitions
increases (\autoref{fig:entropy_ceasers_2048}).
This agrees with the intuition on set associative caches that the leakage correlates with the number of cache sets, which decreases when associativity increases. (Assuming a constant cache size.)
However, caches with 4 or 8 partitions offer better resistance to the eviction-set attack than those with 1 or 16 partitions.

The flexibility of \fw allows us to also compare attack strategies against existing
caches. In particular, we show that the Prime+Prune+Probe approach for eviction set construction
achieves more precise results than the Single Holdout and the Group Elimination Methods.
Moreover, we show that for caches with low randomization, constructing an eviction set is
a good strategy for cryptographic attacks.  However, in highly random designs the cache-occupancy attack presents a more efficient strategy.  Hence we recommend that secure cache designers consider the attack.

\fw is available as an open source project. 
We expect that it can be used for evaluating cache designs by hardware manufacturers and researchers alike.

\section*{Acknowledgments}
This research was supported by
  the Air Force Office of Scientific Research (AFOSR) under award number FA9550-20-1-0425; 
  an ARC Discovery Early Career Researcher Award DE200101577; 
  an ARC Discovery Project number DP210102670; 
  the Blavatnik ICRC at Tel-Aviv University; 
  the National Science Foundation under grant CNS-1954712. 
  the Phoenix HPC service at the University of Adelaide;
  and 
  a gift by Intel. 



\providecommand{\noopsort}[1]{}

\end{document}